# Radiation damage on Silicon Photomultipliers from ionizing and non-ionizing radiation of low-earth orbit operations

Stefano Merzi [a,*], F. Acerbi [a,b], C. Aicardi [c], D. Fiore [d], V. Goiffon [e], A. Gola [a,b], O. Marcelot [e], A. Materne [c], O. Saint-Pe [d]

[a]Fondazione Bruno Kessler (FBK), Center for Sensors and Devices, Via Sommarive 18, Trento, Italy.
[b]TIFPA Trento Institute for Fundamental Physics and Applications, Via Sommarive, 14, Trento, Italy.
[c]Centre National d'Études Spatiales (CNES), 18, avenue Edouard Belin, Toulouse, France.
[d]AIRBUS Defence and Space, 31, Rue des Cosmonautes Toulouse, France.
[e]Institut Supérieur de l'Aéronautique et de l'Espace (ISAE-SUPAERO), 10, avenue Édouard Belin, Toulouse, France.

ABSTRACT

Silicon Photomultipliers (SiPMs) are single photon detectors that gained increasing interest in many applications as an alternative to photomultiplier tubes. In particular in the field of space experiments, where volume, weight and power consumption are a major constraint, their advantages like compactness, ruggedness, and their potential to achieve high quantum efficiency from UV to NIR, makes them ideal candidates for spaceborne, low photon flux detectors. During space missions however, SiPMs are usually exposed to high levels of radiation, both ionizing and non-ionizing, which can deteriorate the performance of these detectors over time.

The goal of this work is to compare process and layout variation of SiPMs in terms of their radiation damage effects to identify the features that helps reducing the deterioration of the performance and develop the next generation of more radiation tolerant detectors. To do this we irradiated with protons and X-rays several NUV-HD SiPMs with small area (single microcell, 0.2x0.2 mm² and 1x1 mm²) produced in FBK. We performed online current-voltage measurements right after each irradiation step and a complete functional characterization before and after irradiation. We compare the results and show the most promising variations for future production of SiPMs for space applications.

## 1. Introduction

Silicon photomultipliers (SiPMs) are solid state photon detectors composed of arrays of passively-quenched Single-Photon Avalanche Diodes (SPADs), also referred to as "microcell", all connected in parallel and working above breakdown condition, in Geiger mode [1]. These detectors have gained increasing interest in the past decade in many fields such as medical imaging [2], automotive LiDAR (Light Detection And Ranging) [3], Cherenkov telescopes [4], and big physics experiments [5]. In recent years the continuous improvement of SiPM performance, combined with their advantages like ruggedness, compactness, lightweight and low operating bias, as well as high photon detection efficiency from the ultraviolet to the near infrared, opened new possibilities for these detectors in space applications, such as satellites operations for cosmic ray studies [6]-[7], atmospheric LiDAR [8], and navigation LiDAR.

When SiPM are used in such kind of applications, as well as in others, like large high-energy physics experiments [9]-[11], the main drawback is the high radiation doses to which they are exposed, deteriorate their performance over time (especially as increase to dark count rate) due to damage caused by both ionizing and non-ionizing radiation effects [12]. In particular ionizing dose induces damages (interface states) at the silicon surface (or interfaces between silicon and dielectrics) while displacement damage dose induces defects in the bulk of the silicon detector. Several studies have been conducted in recent years to evaluate the effects of radiation damage on SiPM functionality [12]-[21]. The majority of the works report that the main effects of radiation damage is an increase in dark count rate (DCR), particularly induced by non-ionizing energy loss (NIEL), proportional to the fluence while minor consensus was found on other parameters, such as signal amplitude, gain and photon detection efficiency.

In FBK we conducted a series of experiments, with protons and X-rays, as reported in [18]-[21], using several SiPM produced over the years with different versions of layout and based on different process technology. While these works provided insightful information on the behaviour of SiPMs after radiation damage, it was not easy to correlate different samples and to isolate the effect of a single parameter in the effects of radiation damage on their performance.

In this paper we present a systematic study of radiation damage on SiPM, of both non-ioniinzng energy loss (NIEL) and ionizing energy loss (IEL). We tested several process and layout splits of FBK SiPMs, based on NUV-HD technology produced specifically for irradiation studies, to easily compare and isolate the effect of a single feature variation on the SiPM response to radiation. The SiPMs were irradiated with protons and X-rays and characterized before, during and after irradiations.

For the proton irradiation level, we chose fluences covering most of the Earth observation and scientific space missions (LEO, GEO, and Lagrange points). The total radiation dose to the detectors varies depending on the type of orbit, on the mission lifetime and on the shielding of the devices in the satellites. A reasonable estimate was calculated for LEO in [17] and [20] with estimated fluences in the order of $10^{11}$ $n_{eq}/cm^2$, which was used as a reference for this study.

On the other hand, regarding ionizing radiation, for low earth orbits it is expected a total dose in the order of $10^2$ Gy/year for a 1 mm Al shield [22]. However previous works showed little to no deterioration of SiPM performance at these doses and, in order to investigate the effect of ionizing radiation on SiPMs, we irradiated the devices up to $10^5$ Gy.

## 2. Devices under test

The devices studied in this work are part of a dedicated SiPM lot, produced in FBK and based on NUV-HD technology [23] with several layout versions and process variations to investigate different aspects of the devices and how they correlate with the radiation-induced performance worsening.

Regarding process splits the focus is on the effects of:

- different electric field profiles and peak values inside the SPADs, to mitigate field enhancement effects on DCR increase after irradiation [24],
- different internal doping profiles in the SPADs,
- different anti reflective coatings (ARCs), particularly the ones used for NUV and VUV sensitive SiPMs, as reported in [23].

The sensors are produced on 3x4 mm² silicon dies, each containing 12 detector structures, divided into three detector sizes (single microcell, 0.2x0.2 mm² and 1x1 mm²) with four microcell sizes each (15, 20, 30 and 40 μm). The focus is on 1x1 mm² devices, while smaller sizes are used after irradiation to perform waveform analysis when the noise is too high to discriminate individual photopeaks on the 1x1 mm² SiPMs. For such SiPMs, we implemented layout splits focused on:

- distance between the active area and the deep-trenches,
- rectangular microcells (instead of the typical square shape) to study the contribution of border effects,
- microcell with metal masking of the dead region outside the active area [26], to investigate possible field plate effects on the charges generated during the irradiation.

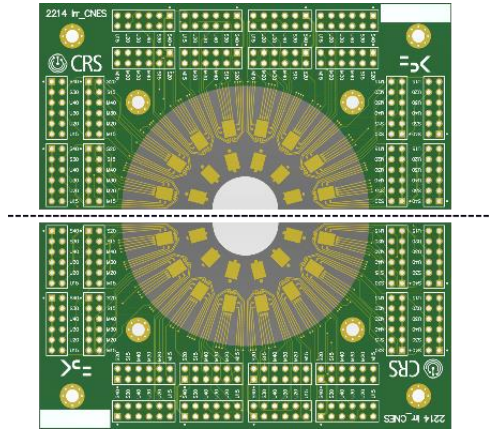

Fig. 1: PCBs used for irradiation tests. The full setup is composed of two PCBs assembled to cover the full irradiation field.

In addition to these devices, we tested also some SiPMs produced with NUV metal-in-trench (NUV-MT) technology, which differentiate from the other ones since the deep trench isolation between microcells are filled with metals instead of just dielectrics, as detailed in [27], with size of 1x1 mm² and single SPADs, with two microcell sizes of 16 μm and 40 μm with metal masking.

To identify the devices, we named them using the wafer number in the production lot, which also defines the process split, and additional characters to define the process split. Naming convention is described in Table 1 and Table 2.

We irradiated some selected chips from each of these wafers, and we characterized them before and after irradiation, with measurement of primary noise (i.e. dark count rate, DCR), gain, optical crosstalk probability (CT), afterpulse probability (AP) and photon detection efficiency (PDE). See [1], [28], [29] for details on the typical characterization procedures. Moreover, during the irradiation campaigns, we performed online measurements of the reverse current-voltage curves (I-V) of the 1x1 mm² devices after each irradiation step, both in dark and under constant illumination conditions.

Table 1: Naming and features of the process splits tested in this work.

| Process Split | Electric field | Afterpulse | ARC |
|---|---|---|---|
| W02 | Low (LF) | Standard | VUV |
| W04 | Low (LF) | Low | VUV |
| W06 | Low (LF) | Standard | NUV |
| W08 | Low (LF) | Low | NUV |
| W16 | Ultra-low (ULF) | Low | NUV |
| W01 (MT) | Low (LF) | Low | NUV |
| W09 (MT) | Ultra-low (ULF) | Low | NUV |

Table 2: Naming and features of the layout splits tested in this work.

| Layout Split | Notes |
|---|---|
| S0 | Standard layout and microcell features |
| D2 | 2-times increased distance between active area and trench |
| AA-DX | X-times (3, 5, 8) increased distance between active area and trench |
| M* | Metal mask outside the active area |
| TS | Rectangular cell: 30 μm 1-6 aspect ratio, 40 μm 1-12 aspect ratio |

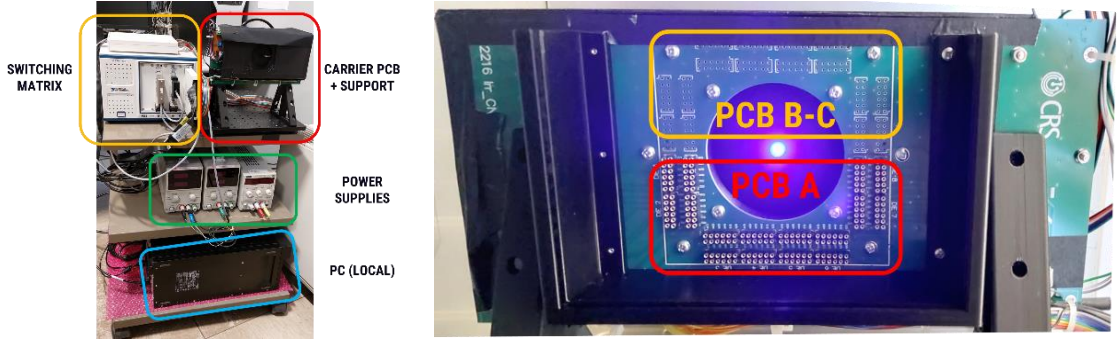

Fig. 2: Left: complete irradiation setup with the four main parts highlighted. Right: detail of the carrier PCB with motorized shutter and blue LED, with highlighted the positioning of the two PCBs with the samples to be irradiated. Note: the system is seen here from the backside. The LED is mounted on the shutter and, when closed, illuminates the SiPMs from the front, i.e. same direction of the incoming proton or X-ray beam.

## 3. Experimental Setup

Tested devices have been mounted on a custom PCB, designed to readout up to 96 channels from 8 die sites, with 12 bonded devices per die. As it can be seen in Fig. 1 the dies in one of the PCB are placed on a half-circle, whose diameter have been chosen to be fully included in the estimated uniform-fluence region, as well as uniform X-ray dose region, at the two irradiation facilities respectively.

Two of these PCBs are used simultaneously during the irradiation (i.e. mounted together on the support, and exposed simultaneously during irradiation), one placed below the other and rotated at 180 degrees. In this way it is possible to use the full uniform fluence region and double the number of irradiated devices. The first PCB undergoes all irradiation steps, IVs of 32 channels (1x1 mm² SiPMs) are measured after each irradiation steps. Conversely, in the second slot, the PCBs are not connected to the readout for online measurements, but they will be tested offline after the irradiations. During the irradiation campaign, several copies of these PCBs are packaged and irradiated at different dose levels. In this way we were able to have both online measurements from the same samples at all fluences (doses), as in the irradiation described in [21], as well as the possibility to have dies irradiated at several fluencies (doses) for offline full characterization, as in the irradiation campaign described in [18].
The complete setup for irradiation is composed of four main parts, shown in Fig. 2 on the left:

- Carrier board for the PCBs to be irradiated.
- Black box to ensure darkness during IV measurement.
- Motorized stage, open during irradiation to avoid partial stopping of the beam, closed to ensure darkness during IV measurement.
- Blue LED for IV measurement under illumination.
- Measurements system, composed by a National-Instrument Switching matrix with 32 channels in a 4x8 configuration (four channels measured simultaneously) and source-and-measurements unit.
- Power supplies for controlling the LED and the motorized stages.
- Local computer for the control of the instrumentation and remote computer for user control of the setup during the irradiation.

Fig. 2 on the right shows the inside of the black box on the carrier PCB with the position of the two PCBs to be irradiated. The primary PCB, measured online, is at the bottom and it is connected to the headers for the 96 channels, while the secondary PCBs are placed at the top and can be easily replaced between irradiation steps.

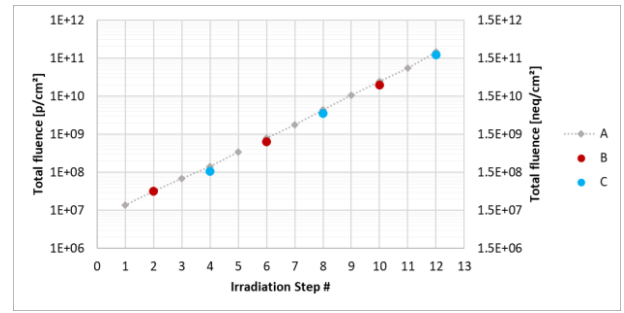

Fig. 3: Total fluence as a function of the irradiation steps for the different PCBs.

## 4. Irradiation facilities and steps

### 4.1. Proton irradiation

The proton irradiation campaign was held at the proton therapy facility in Trento [30] in April 2023. The irradiation took place in the research line using the "dual ring" setup, with uniform beam intensity over a circle with a diameter of 60 mm. The proton beam is emitted at an energy of 148 MeV, which was lowered down to 74 MeV using 10 panels of RW3 [31] (1 cm thick). Proton beam current can be tuned between 1 nA and 300 nA . The total amount of charge, proportional to the number of protons generated during each irradiation step, is measured through a Lynx ionization chamber [32]. This was also used before irradiation to verify the shape and uniformity of the beam. The proton flux in this configuration was measured at $3.84 \cdot 10^5$ p·s$^{-1}$·cm$^{-2}$·nA$^{-1}$.

Irradiation of the devices was carried out in 12 steps for two days. Each step has an exponentially increasing fluence compared to the previous one to span a wide range of fluences with exponentially spaced sampling. The primary PCB, placed on the bottom and measured online, is called PCB-A. Secondary PCBs are named PCB-B-X and PCB-C-X where X is a progressive number of each copy of the PCB. Each secondary PCB is irradiated two steps (of the PCB-A) and then replaced with a different one. Fig. 3 shows the total fluence as a function of the irradiation steps for PCB-A and for the secondary PCBs. As can be seen, in the semi-logarithmic scale the points are evenly spaced in fluence with a factor of 2.3 between each step.

Unfortunately, due to a malfunctioning of the measurement setup during the two highest fluence irradiation steps, the IV curves just after irradiation have not been measured correctly. It was possible to extrapolate the data of the last point (highest fluence) by interpolating annealing data of current vs time with an exponential and projecting back the current at t = 0. However, due to possible inaccuracy in such procedure, these data have been used for the visual indication on the plots, but not used for more quantitative analysis.

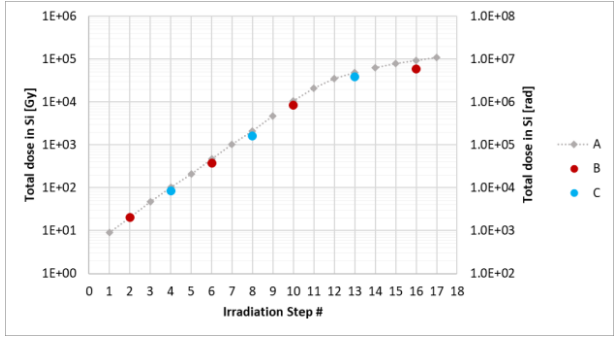

Fig. 4: Bottom: Total dose in silicon as a function of the irradiation steps for the different PCBs.

### 4.2. X-rays irradiation

The X-ray irradiation campaign was held at the TIFPA (Trento Institute for Fundamental Physics and Applications) in Trento between 22/05/2023 and 24/05/2023. The irradiation used an X-ray tube with tungsten anode, biased at 40 kV, with 180um Aluminium filter in front. It produced a 33° uniform dose region with main lines between 7.6 keV and 12 keV and Compton emission up to 40 keV. The dose rate was adjusted by changing the current, between 10 mA and 25 mA and it was measured using an ionization chamber (PTW 30010 Farmer). This instrument was previously calibrated (conversion factor) to have the precise dose in silicon (the one reported in the paper). The SiPM tested were placed 20 cm from the tube. The dose rate in this configuration was measured to be 12.5 Gy·s⁻¹·mA⁻¹.

Irradiation of the devices was carried out in 17 steps during three days. As for the proton irradiation, each step has an exponentially increasing fluence (excluding the last steps that followed a linear trend) compared to the previous one to span a wide range of fluences while maintaining an exponentially spaced sampling. As for the proton irradiation the bottom PCB, called PCB-A, was fixed and measured online, whereas PCB-B-X and PCB-C-X (top) were irradiated only for a few steps and then replaced with a different one. Fig. 4 shows the total dose in silicon as a function of the irradiation steps for the main PCB and for the secondary ones.

## 5. Results of proton irradiation

### 5.1. Breakdown voltage and reverse current

From the online measurements, specifically IV curves measured under illumination, we first extrapolated the breakdown voltage ($V_{BD}$) using the second logarithmic derivative method (as used and reported in [18]) and we compared the current in dark condition both at 5 V below breakdown (called "non-multiplied current") and 3 V above breakdown (i.e. 3V of "excess bias") respectively. The current above breakdown is called "multiplied current".

Fig. 5 at the top shows that the breakdown voltage does not change with fluence (up to the maximum level reported) for all layout and process splits.

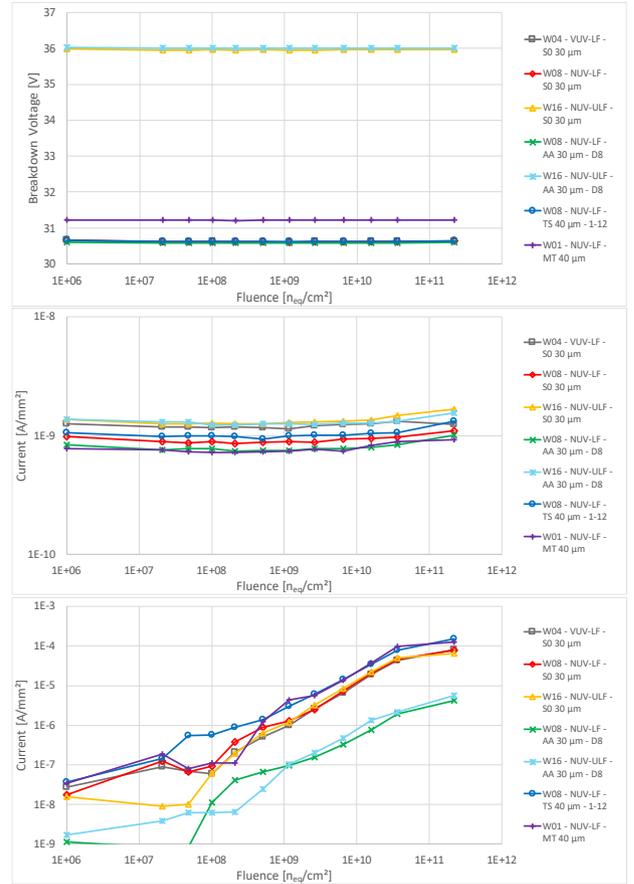

Fig. 5: Top to bottom: breakdown voltage, non-multiplied current (5 V below breakdown) and multiplied current (3 V above breakdown) as a function of the proton fluence for different structures of PCB-A.

Devices with breakdown voltage around 36 V are ULF technologies, while LF technologies have breakdown voltage around 31 V. Also, the current below breakdown, shown in Fig. 5 in the middle, do not show significant variation. Small variations are visible at the highest fluences.

Fig. 5 bottom shows the current measured 3 V above the breakdown (multiplied current). For fluences up to $4 \cdot 10^7$ $n_{eq}/cm^2$ there is mostly no change in multiplied current. Some devices show a random increase in current with many of them being annealed between irradiation steps. For fluences above $1 \cdot 10^9$ $n_{eq}/cm^2$ it is observed a linear increase of current with dose with no major difference in behaviour between devices. Between these two fluences there is a "transition region" with mostly linear increase of current with fluence and with strong variation between samples and irradiation steps. The main trend observed between samples is that devices with smaller fill factor (FF, i.e. the ratio between the active area and the total area of the microcell), have smaller current than devices with larger FF due to the reduction in the high field volume, possibly further reduced by border effects, which in turns reduces the DCR, which is likely mainly due to field-enhanced SRH generation [33].

### 5.2. Dark count rate

Different cell sizes have different gain values, so to remove this contribution from the comparison it is calculated the dark count rate for the different SiPMs as a function of the fluence. DCR is calculated from the

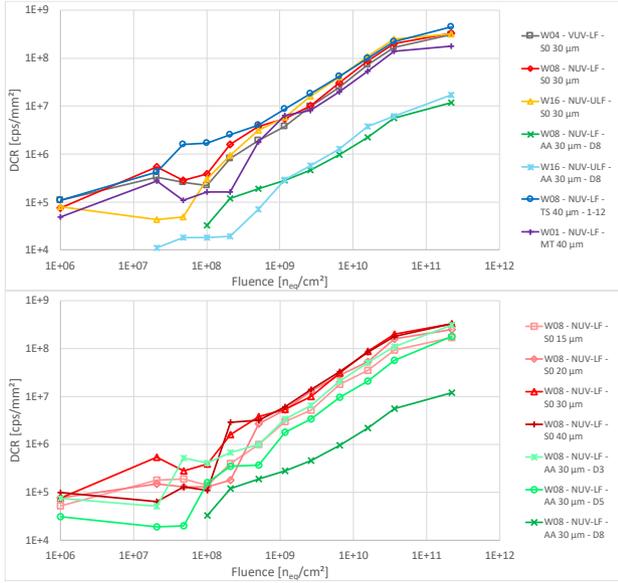

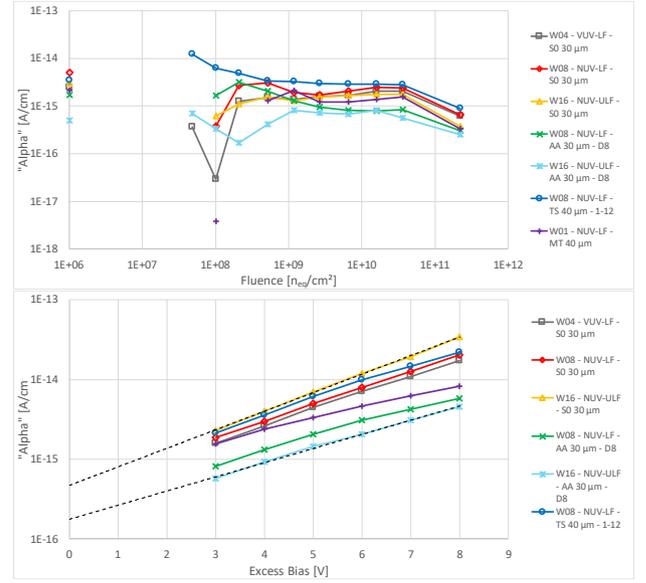

Fig. 6: Calculated DCR as a function of the proton fluence for different structures of PCB-A, calculated at 3 V of excess bias. The bottom plot highlights the difference in DCR between cells with different size and different FF.

Fig. 7: Damage parameter as a function of the proton fluence (top) at 3 V of excess bias and as a function of the excess bias at $2.4 \cdot 10^{10}$ p/cm$^2$ (bottom).

multiplied current knowing the gain (G) and the Excess Charge Factor (ECF) according to the formula:

$$DCR = \frac{I}{q * G * ECF}$$

Where q is the elementary charge. Under the assumption that the gain and the ECF do not change with irradiation (which was verified in [21]) it is possible to calculate the product of gain and ECF, also known as Gain Current (GC) from current and DCR measurements of non-irradiated devices, and use this value to estimate the DCR after irradiation, when direct measurement through waveform analysis is no longer possible due to the high noise of the detectors.

Fig. 6 shows the calculated DCR as a function of the proton fluence for the different structures. Its behaviour is similar to what showed by the multiplied current, with a lower DCR for structures with very low active area (AA 15µm-D3, AA 30µm-D8) compared to other structures. This is related to the lower fill factor of these structures, and consequently, to the size of the high field region compared to the cell size: for the same defect density, there is a lower total number of defects in the high field region of low AA structures compared to other structures with the same cell size, resulting in a lower value of DCR. For the same reason it is observed a direct relationship between cell size and DCR, where smaller cells sizes, due to the smaller FF also show a smaller DCR (Fig. 6, bottom).

### 5.3. Damage factor

To take into account the presence of gain and excess charge factor in a SiPM, the damage factor α is calculated for each fluence point according to the formula:

$$\alpha = \frac{\Delta DCR * q}{\Phi * V * P_t} = \frac{\Delta I}{\Phi * V * P_t * (Gain * ECF)}$$

Where Φ is the proton fluence and V is the volume of the high field region and $P_t$ is the triggering probability of the avalanche reported in the work by [34]. This term is necessary to take into account that not all damages generated by a proton result in an avalanche, and this avoids underestimating the damage parameter, especially at low bias when the triggering probability is low.

Fig. 7 at the top, shows the calculated damage factor for all devices as a function of the fluence at 3 V of excess bias. For fluences below $8 \cdot 10^8$ p/cm$^2$ there is a high uncertainty in the calculation due to the "random" behaviour of DCR at low fluences explained in the previous section. Above this value the damage factor is constant with fluence, excluding the highest point in which there might be either saturation of the detectors or errors in the reconstruction of its pre-annealing currents. All devices show a damage factor of around $10^{-15}$ A/cm and differences between devices are rather small, as in the case of DCR normalized by the FF. This is due to the presence of the factor V, the high-field region volume, calculated from the FF of the microcell and the thickness of high-field region (neglecting field effects at the border of the active area or in the low-field depleted region). The lack of major differences in the damage factor between different devices suggests that the damage factor, in first approximation, is not easily modified with process or layout splits and improvements in the performance of SiPM after irradiation might need to focus on reduction of the high-field region in combination with light focusing (e.g. microlens) or charge focusing mechanisms to increase the PDE of the detector while reducing the radiation-sensitive area.

Fig. 7 at the bottom, shows the damage factor calculated at different excess bias for the highest fluence with direct measurements (not reconstructed). It shows higher values ($10^{-15}$ A/cm) compared to the one expected for detectors without gain ($3.5 \cdot 10^{-17}$ A/cm) and an increase in damage factor with the excess bias. This indicates the presence of field-enhancement effects on the generation rate of the carriers. Extrapolating the damage parameters at lower excess biases, so a lower value of electric field, these values approach the ones reported in literature, ranging from $2 \cdot 10^{-16}$ and $7 \cdot 10^{-16}$ A/cm.

### 5.4. Photon Detection Efficiency

In Fig. 8 it is reported the measured and estimated PDE at 3 V of excess bias for a wavelength of 420nm for the 32 samples under investigation.

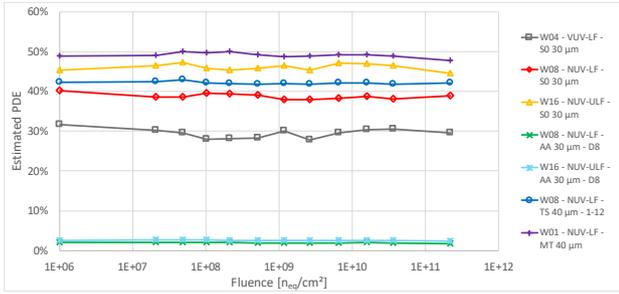

Fig. 8: PDE as a function of the proton fluence for the 32 structures of PCB-A, calculated at 3 V of excess bias.

Samples from S0 dies have an expected PDE between 31% and 53% depending on the SPAD size. Cells with increased distance between trench and active area have lower FF and lower PDE. For the 30 μm cells, the PDE of S0 is 40%, which reduces to 34% in the D2 version and to 27%, 13% and 4% for the AA die, with D3, D5, and D8 version respectively.

The PDE measured before irradiation is used to calculate the behaviour of PDE after irradiation by using the detector response measured through the IV curves taken under illumination at different fluences. There are some fluctuations in PDE as a function of the fluence but there seems to be no significant change in PDE, even at high fluences, excluding the highest point, which is affected by the issues discussed above. The only minor exception is given by the VUV samples that show a decrease in relative PDE of ~10% at low fluences but they have a behaviour similar to the other samples at higher fluences.

### 5.5. Gain, Crosstalk and Afterpulse

In Fig. 9 are reported the gain, crosstalk probability and afterpulse probability measured at 4 V of excess bias before and after irradiation. Gain and AP were measured on SPADs at room temperature, CT measurement was performed on 1x1 mm² at room temperature and on 0.2x0.2 mm² SiPMs at -20°C after irradiation, to discriminate single events in high DCR conditions.

No significant change in gain is observed after irradiation, apart from some fluctuation caused by the uncertainty of the measurement on samples with high noise level. For CT measurements, time constraints and instability of the amplifiers operated at low temperatures limited the number of tested devices, but the results do not show a link between radiation damage and a change in crosstalk probability.

In the case of AP measurements, it is observed a significant increase for almost all devices after irradiation. This effect is related to the defects generated in the silicon by NIEL. The defects act as a trapping centres for charges, thus increasing the AP probability for irradiated samples. A few samples do not show this increase, but this is mainly attributed to limitations in the measurement setup and software, which is not able to detect AP events below a certain amplitude and will underestimate their number in case of cells with long recharge time.

Despite the increase in AP, its contribution to the total ECF is limited as its probability remains relatively low. Moreover, the gain and the CT do not show a significant variation in samples measured before and after irradiation. This proves the validity of the hypothesis used in the calculation of the DCR, where it was assumed that the Gain current, i.e. the gain multiplied by the excess charge factor, do not change with proton irradiation.

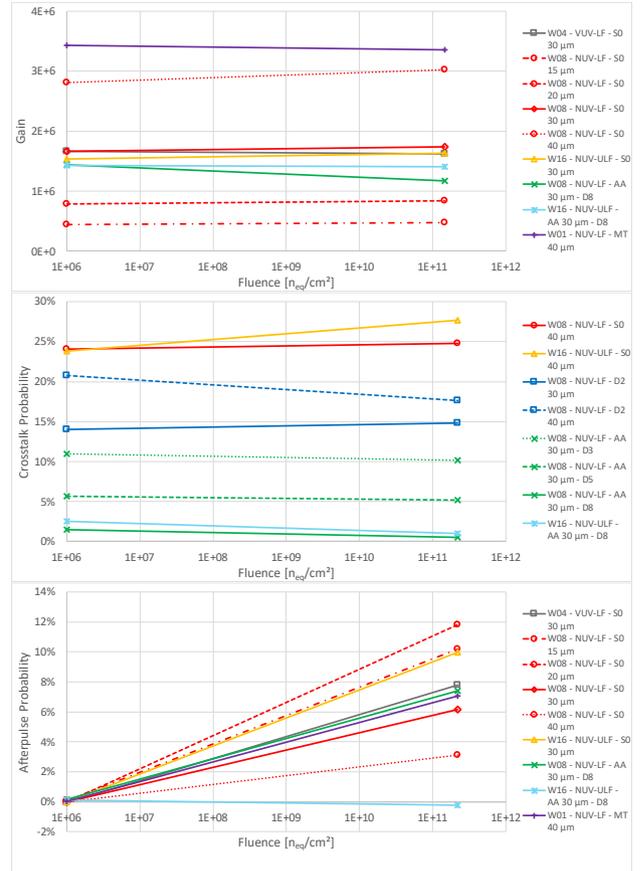

Fig. 9: Top to bottom: gain, crosstalk probability and afterpulse probability measured before and after proton irradiation at 4 V of excess bias.

### 5.6. Activation energy

To calculate the activation energy (Ea) of the SiPMs it was taken the IV measurements at different temperatures. At the same time, it was measured also the temperature dependence of the breakdown voltage with temperature. For irradiated samples the IVs were taken at temperatures between -30°C and 20, with 2°C steps, to avoid annealing of the samples. For non-irradiated samples on the other hand, the IV measurements were taken at temperatures between 20°C and 40°C, to have higher currents and allow a better analysis of the data. Because of the high variability in Ea data, due to uncertainty in the current measurements for devices with low gain, we averaged this measurement over the four structures of each die to improve readability of the results.

Fig. 10 shows the calculated activation energy for the 32 structures before and after irradiation. Before irradiation the samples show an activation energy between 0.6 and 0.9 eV with no clear correlation between process splits. For irradiated samples LF structures have a lower Ea, around 0.37 eV while ULF structures have a higher activation energy, around 0.4 eV. This is reflected in the temperature difference needed to halve the dark current (and DCR). Before irradiation the halving temperature is around 7°C while after irradiation it is ~12.3°C for LF structures, and it is reduced to ~11.5°C for ULF structures.

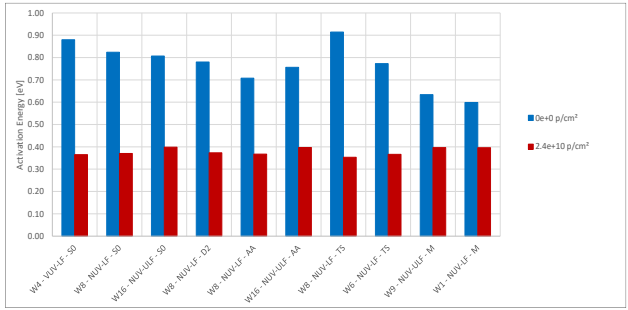

Fig. 10: Average activation energy for the different dies calculated at 3 V of excess bias before and after proton irradiation.

# 6. Results of X-ray irradiation

## 6.1. Breakdown voltage and current

Fig. 11 at the top shows the breakdown voltage as a function of the total dose in silicon for the different structures. As in the case of proton irradiation, the breakdown voltage does not change with fluence for all layout and process splits. Small variations in breakdown, observed for all devices, are likely related to small temperature changes of the irradiation room, with a lower temperature in the morning that increases during the day.

Fig. 11 in the middle shows the leakage current measured 5 V below the breakdown. No change is observed up to few hundreds of Gray. For higher doses it is observed a more than linear increase in current for all devices with the faster increase measured for W6-S0 and W16-S0 devices (NUV-LF and NUV-ULF).

Fig. 11 at the bottom shows the dark current, measured 3 V above the breakdown. As in the previous case, no change is observed up to 1000 Gy. For higher doses it is observed a fast increase in dark current for all NUV (W6, W8, W16) devices excluding the one with the metal masking (M*): W16 ULF shows the highest current that reaches saturation of the detector while W6-AA shows lower currents due to the small active area of the detector. On the other hand, VUV devices (W2, W4) and devices with metal masking (W6-M*) show a much slower increase in dark current with the dose.

## 6.2. Dark Count Rate

Dark count rate was calculated using the same method described above. Fig. 12 show the DCR as a function of the total dose for the different structures. All devices show a DCR around or below $10^5$ c.p.s./mm² at room temperature before irradiation. Lower DCR is observed for devices with small AA, due to the smaller FF.

After irradiation devices from the VUV process split and NUV samples with metal masking show a relatively low DCR with less than an order of magnitude increase, below $10^6$ c.p.s./mm². The opposite behaviour is observed for NUV-ULF samples, with saturation of the detectors above $5 \cdot 10^4$ Gy, and DCR in excess of $10^9$ c.p.s./mm². NUV samples without metal masking show a higher DCR, that decreases with the reduction of the active area and the increase of the distance between the trench and the high field region. While partially hidden by the sample-to-sample variability, it is present an inverse relation between cell size and DCR with smaller cells showing the higher DCR after irradiation (Fig. 12, bottom). This is the opposite of what is observed in the case of proton irradiation, and it is likely due to charges accumulated in the insulating oxide in the trenches: smaller cells have longer trench perimeter per unit area and this can result in a higher DCR.

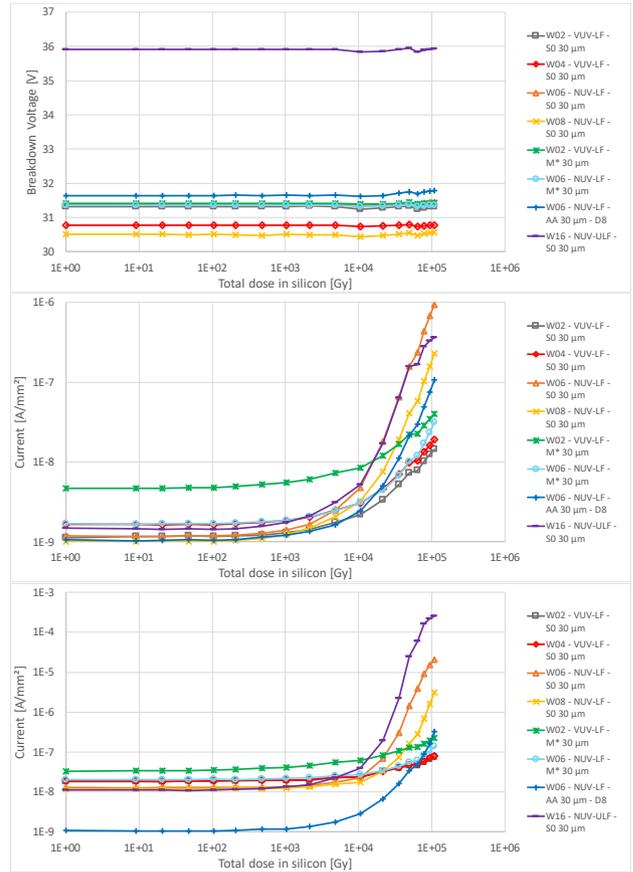

Fig. 11: Top to bottom: breakdown voltage, non-multiplied current (5 V below breakdown) and multiplied current (3 V above breakdown) as a function of the total dose in silicon for the 32 structures of PCB-A irradiated with X-rays.

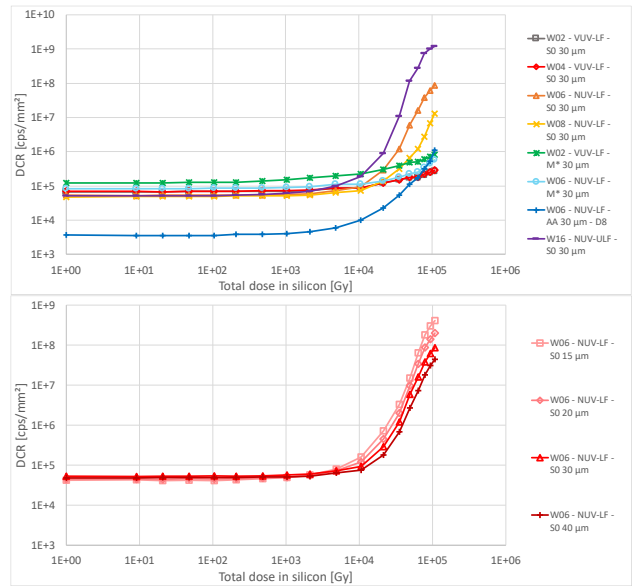

Fig. 12: Calculated DCR as a function of the X-ray dose for the 32 structures of PCB-A, calculated at 3 V of excess bias. The bottom plot highlights the difference in DCR between cells with different sizes.

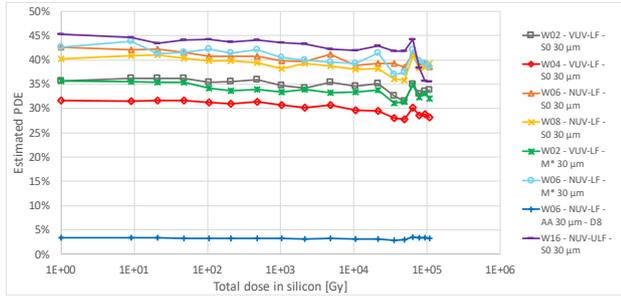

Fig. 13: PDE as a function of the X-ray dose for the 32 structures of PCB-A, calculated at 3 V of excess bias.

### 6.3. Photon Detection Efficiency

Fig. 13 shows the estimated PDE at 420nm as a function of the X-ray dose. A few outliers, mostly from the NUV-ULF and NUV-LF samples, show a strong reduction in PDE above $10^4$ Gy likely due to the high DCR that causes saturation of the detectors at high doses. For all other samples it is observed a small reduction in PDE with the dose, in the order of 10% relative to the initial value, for all structures independent of the technology or layout split. A possible explanation can be the accumulation of fixed charge in the oxide layer at the surface of the detector, which can alter the electric field of the active region and reduce the PDE. Charge accumulation in the trenches, on the other hand, does not seem to affect the PDE, as there is no difference in behaviour between samples with different distance between trench and high field region. Further investigation needs to be done to exclude possible setup issues, such as instability or degradation of the LED used for the measurement.

### 6.4. Gain, Crosstalk and Afterpulse

In Fig. 14 are reported the gain, crosstalk probability and afterpulse probability measured at 4 V of excess bias before and after irradiation. The limited time availability before irradiation did not allow to completely characterize the samples before irradiation and their parameters were estimated from measurements performed on similar devices (represented with dashed lines in the plots) while the parameters after irradiation were measured directly. In general, it is observed no significant change in the parameters with the dose but it was observed an increased variability in their value after irradiation. This is mainly attributed to the high electronic noise of the setup, which worsen the single photoelectron detection capability and, consequently, the calculation of these three parameters. One exception is the reduction of gain observed for all layout splits of W06-AA, with 30 μm cells with reduced active area. It is possible that the accumulated charge in the trench oxide changed the shape of the electric field and the depletion region, reducing the gain of the cells but more studies are needed to confirm this effect. A second exception is given by the increased AP probability of W02-M0*, devices with metal masking outside the active area, which show an absolute increase in AP between 2% and 5% for all layout splits. These devices are from the VUV-LF process split with standard AP surface configuration and it is possible that the presence of the metal masking for in VUV structures might increase the sensitivity of the detector to surface TID damage which results in an increased AP probability. Also in this case more studies need to be performed to confirm the results and better understand the effect of the metal masking on the device surface.

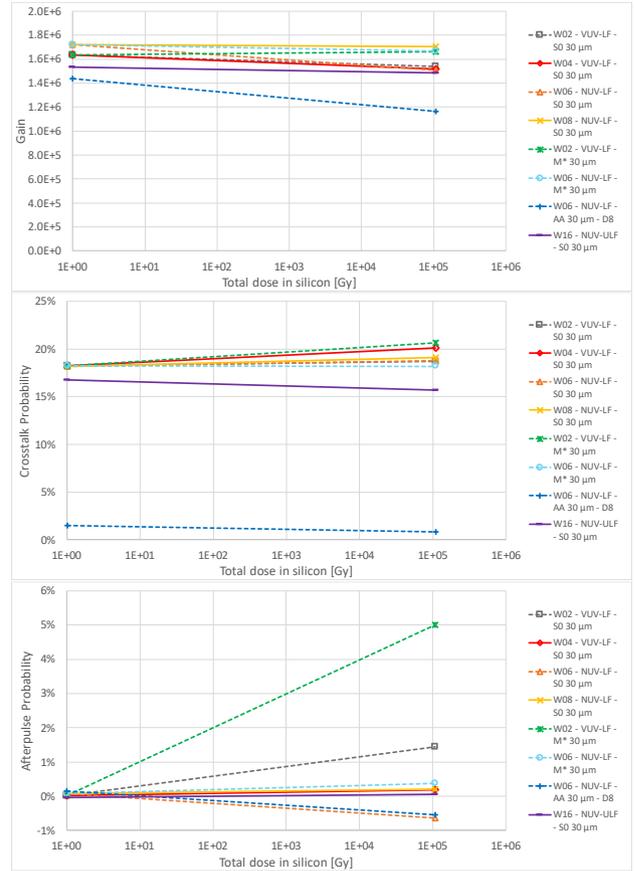

Fig. 14: Top to bottom: gain, crosstalk probability and afterpulse probability measured before and after X-ray irradiation at 4 V of excess bias.

As in the case of protons, these measurements prove the validity of the initial hypothesis used in the calculation of the DCR, where it was assumed that the Gain current do not change with irradiation.

### 7. Conclusions

We irradiated 32 structures from VUV-LF, NUV-LF and NUV-ULF technology (different ARC and electric field configuration), with S0, D2, AA and TS layouts (increased distance between trench and active area and increased perimeter of the high field region) with 74 MeV protons up to $1 \cdot 10^{11}$ p/cm$^2$, or $1.5 \cdot 10^{11}$ 1 MeV $n_{eq}$/cm$^2$. No change in breakdown voltage, non-multiplied current or PDE is observed up to the highest fluence. A linear increase in multiplied current and DCR is measured starting at ~$10^8$ p/cm$^2$ for all devices without major difference in damage factor between different process and layout splits: it is possible that uncertainties in the delivered dose and in the measurements might hide differences between process splits, in particular between LF and ULF. The main difference in behaviour after irradiation is observed for AA structures, where the small active area results in a lower value of DCR after irradiation, proportional to the microcell fill factor. Irradiated devices showed a lower activation energy compared to before irradiation. In this case ULF, thanks to their lower peak electric field, showed a higher activation energy compared to LF structures, resulting in a faster reduction of DCR with cooling.

We irradiated 32 structures from VUV-LF, NUV-LF and NUV-ULF with standard and low AP (different ARC, electric field, and surface doping configuration) with S0, AA and M0* layouts (increased distance between trench and active area and masking of the area outside the high field region with a metal layer) with X-rays from a Tungsten target at 40 kV up to $10^5$ Gy. No change in breakdown voltage is observed up to the highest dose. An increase in both non-multiplied and multiplied current with dose is measured for all structures, with the larger increase observed for NUV-LF and NUV-ULF structures, while VUV-LF structures and NUV-LF-M0* structures showed a smaller increase. First analysis suggests that smaller cells show a larger increase in DCR, probably related to the higher trench perimeter per unit area. Compared to proton irradiation, multiplied current and DCR does not increase as much at most space mission orbits X-rays dose level, thus showing that TID is not the main contributor to DCR when compared to displacement damage dose. A small reduction in PDE as a function of the dose is observed for all devices, possibly due to charge accumulation in the oxide layer.